\documentclass[aps,prx,onecolumn,notitlepage,nofootinbib,longbibliography]{revtex4-2}

\usepackage[T1]{fontenc}
\usepackage[utf8]{inputenc}
\usepackage{lmodern}
\usepackage{microtype}
\usepackage{amsmath,amssymb}
\providecommand{\braket}[1]{\ensuremath{\langle #1 \rangle}}
\usepackage{booktabs}
\usepackage{float}
\usepackage{graphicx}
\usepackage{xcolor}
\usepackage{listings}
\usepackage{tikz}
\usetikzlibrary{positioning,arrows.meta}
\usepackage{pgfplots}
\pgfplotsset{compat=1.18}
\usepackage[colorlinks=true,linkcolor=blue!50!black,citecolor=blue!50!black,urlcolor=blue!50!black]{hyperref}

\definecolor{codebg}{RGB}{247,247,249}
\definecolor{codekw}{RGB}{0,92,153}
\definecolor{codestr}{RGB}{163,21,21}
\definecolor{codecomment}{RGB}{194,86,8}
\newcommand{\qarp}{OpenQARP}
\newcommand{\ms}{\,ms}
\newcommand{\dg}{\textsuperscript{\dag}}

\begin{document}

\title{OpenQARP: a modular framework for quantum application research}

\author{Stefano Scali}
\author{Vicente P. Soloviev}
\author{Antonio M\'arquez Romero}
\author{Brian Coyle}
\author{Giuseppe Buonaiuto}
\author{Annie Paine}
\author{Jonathan H. Fetherolf}
\author{Marcos Diez Garc\'ia}
\author{Michal Krompiec}
\author{Josh Kirsopp}
\affiliation{Fujitsu Research of Europe}


\begin{abstract}
We introduce \qarp{}, the Open Quantum Application Research Package: an
open-source Python framework for quantum application research, built on a
compiled C++ core. In \qarp{}, an application is assembled from an algorithm library that
orchestrates three interchangeable layers: \emph{blocks} describe circuits,
\emph{primitives} describe what to extract from them, and \emph{engines}
describe how they run. The release spans near-term through fault-tolerant
methods, targeting problems from electronic structure to combinatorial
optimization, with device-aware compilation, circuit cutting, and resource
estimation over a single set of numerical conventions that the code must
follow. Every numerical feature is checked against an independent oracle, and
every published timing carries a correctness check. Against OpenFermion, Qiskit/Aer, PennyLane/Lightning, Qulacs, qsim, and
pytket, \qarp{} is one to two orders of magnitude faster on operator algebra
and among the fastest on simulation. It matches mature compilers on two-qubit
gate count, and expresses a complete algorithm in the lines a framework needs
rather than the plumbing a primitive stack demands.
\end{abstract}

\maketitle

\section{Introduction}
\label{sec:intro}

Quantum application research lives in software. Before any experiment reaches
hardware, a researcher has to express the algorithm as circuits, attach
the right measurement scheme, simulate at a meaningful scale, compile to a
device's constraints, and estimate what the complete computation would cost. Each of
these steps must also be numerically trustworthy, because convention mistakes
in quantum software rarely cause programs to crash; they produce plausible,
wrong physics.
This paper presents \qarp{}~\cite{openqarp2026}, an open-source framework
built to make that whole workflow fast and trustworthy at once.

The existing ecosystem already offers excellent tools for individual steps:
circuit-centric software development kits (SDKs) such as
Qiskit~\cite{javadiabhari2024qiskit} and Cirq~\cite{cirq2024}, differentiable
quantum workflows in PennyLane~\cite{bergholm2018pennylane}, compilers
such as pytket~\cite{sivarajah2020tket}, high-performance simulators such as
Qulacs~\cite{suzuki2021qulacs} and qsim~\cite{qsim2020}, and operator layers
such as OpenFermion~\cite{mcclean2020openfermion}. In practice a researcher
combines several of these, and the combination is where research time is
lost: every package brings its own qubit ordering, its own phase
conventions, its own data model, and its own performance pitfalls.

\qarp{}, developed at Fujitsu Research of Europe and now released as open
source~\cite{openqarp2026}, is our answer to that integration problem. It was
built for the algorithm researcher first, then grown outward: toward the
hardware researcher, with devices, routing, noise, and resource estimation,
and toward the non-specialist, with algorithms that run end to end rather than
needing to be assembled. Throughout, the physics stays the user's concern and
the C++ engine underneath does not. The framework is also built for a field
that keeps changing. Algorithms, simulators, and compilation techniques that
are state of the art today will be superseded, so \qarp{} is organized around
\emph{interfaces} rather than around any particular method. Its design rests on three
commitments:

\begin{enumerate}
\item \textbf{A compositional model.} Circuits are \emph{blocks}, the
  quantity of interest is a \emph{primitive}, and execution is an \emph{engine}.
  Everything else in the framework, from variational loops to phase
  estimation to circuit cutting, is these three pieces with more interesting
  parts plugged in. Each axis is open: users and maintainers add blocks,
  primitives, and engines without touching the others.
\item \textbf{One set of conventions.} A single authoritative document fixes
  endianness, angle conventions, and phase behavior for the whole codebase,
  so components written years apart still compose correctly.
\item \textbf{Correctness as evidence.} Every numerical feature is tested
  against an independent reference, and every published benchmark number
  carries a correctness check. Performance claims that cannot be re-verified
  by a third party are not published.
\end{enumerate}

A framework is usually expected to cost performance against a script written
straight onto a simulator kernel. Section~\ref{sec:benchmarks} measures the
reverse. Because a block says what a circuit does and an engine says how it
runs, the engine is handed the complete gate list and the complete sum of
Pauli terms before any of it executes. It can therefore fuse neighboring
gates into dense multi-qubit blocks and contract every term sharing a flip
mask in one pass over the state, while a script that submits one gate and one
term at a time can do neither. The separation
that makes the framework extensible is also what makes it fast.

Pre-release versions of \qarp{} produced the numerical results in several of
our own publications, and several of the methods surveyed in
Section~\ref{sec:algorithms} were developed in them: a quantum computing
quantum Monte Carlo framework built on structured state
preparation~\cite{buonaiuto2026qcqmc}, a seniority-zero state-preparation
method whose UpCCD ansatz ships in the block
library~\cite{krompiec2026seniority}, a study of nuclear magnetic resonance
spectra with MMQCELS~\cite{marquez2026nmr}, and a purified phase-estimation
protocol~\cite{scali2025dosqpe}.

\qarp{} did not begin as a framework. It started as internal research code,
written by hand in Python for one project at a time. The rewrite that produced
the current architecture was forced by performance: the cost of crossing the
Python/C++ boundary dominated every workload that re-executes one circuit at
many parameter settings, an optimizer sweeping an ansatz and a
phase-estimation ladder repeating a Trotterized evolution alike, which is why
the compiled core now holds gate parameters symbolically and binds them at run
time instead of rebuilding a circuit on every call
(Section~\ref{sec:core}). Since that rewrite the framework has been developed
with the help of coding agents, under the conventions and gates described in
Section~\ref{sec:agents}.

The rest of the paper describes the design (Section~\ref{sec:philosophy}),
surveys the current feature set (Section~\ref{sec:features}), explains the
testing methodology (Section~\ref{sec:testing}), presents benchmarks against
the established ecosystem (Section~\ref{sec:benchmarks}), and describes how
the framework itself is developed (Section~\ref{sec:agents}).

\section{Design}
\label{sec:philosophy}

\begin{figure}[t]
\centering
\begin{tikzpicture}[
  font=\small,
  layer/.style={draw, rounded corners=2pt, minimum height=2.3cm, text width=4.0cm,
    align=center, fill=#1},
  side/.style={draw, rounded corners=2pt, text width=4.0cm, align=center,
    fill=gray!8, font=\footnotesize, minimum height=1.3cm},
  bar/.style={draw, rounded corners=2pt, text width=13.2cm, align=center,
    fill=gray!8, font=\footnotesize, minimum height=0.7cm},
  arr/.style={-{Latex[length=2mm]}, thick},
  darr/.style={{Latex[length=2mm]}-{Latex[length=2mm]}, thick},
]
\node[layer=blue!8] (blocks) at (0,0) {\textbf{Blocks}\\ what the circuit does\\[3pt]
  {\footnotesize \texttt{SimpleBlock}, \texttt{CompositeBlock}, \texttt{ControlledBlock};
  $\sim$70 provided: ans\"atze, state preparation, QFT, QPE, Trotter, LCU, QSP/QSVT, \ldots}};
\node[layer=green!8, right=0.6cm of blocks] (prims) {\textbf{Primitives}\\ what to extract\\[3pt]
  {\footnotesize \texttt{StateVector}, \texttt{Sampler}, \texttt{PauliAveraging},
  classical shadows, Hadamard and SWAP tests, \ldots}};
\node[layer=orange!10, right=0.6cm of prims] (engines) {\textbf{Engines}\\ how it runs\\[3pt]
  {\footnotesize CPU statevector, GPU;
  compile: rebase, optimize, route}};
\draw[arr] (blocks) -- (prims);
\draw[arr] (prims) -- (engines);
\node[bar, above=0.5cm of prims] (algos) {\textbf{Composite algorithms}: VQE, ADAPT,
  VQD, SSVQE, QSE, QPE, \mbox{DOS-QPE}, QMEGS, MMQCELS, QAOA, PCE, VFF, QITE, Monte Carlo,
  Grover, Shor, amplitude estimation, circuit cutting, \ldots \quad$+$\quad \textbf{classical optimizers}};
\draw[arr] (algos) -- (prims);
\node[side, below=0.5cm of blocks] (interop) {\textbf{Interop}\\ emit and absorb: Qiskit,
  PennyLane, pytket, Qulacs, OpenQASM, \ldots};
\node[side, below=0.5cm of prims] (resources) {\textbf{Resources}\\ counts at every
  compilation stage, Clifford$+T$ synthesis, \ldots};
\node[side, below=0.5cm of engines] (device) {\textbf{Device} (data)\\ qubit count, coupling
  map, gate set, noise model, \ldots};
\draw[darr] (interop) -- (blocks);
\draw[arr] (device) -- (engines);
\draw[arr] (engines) -- (resources);
\end{tikzpicture}
\caption{Architecture. A block says what the circuit does, a primitive says
what to extract from it, and an engine says how it runs; composite algorithms
are orchestrations of primitives and classical optimizers. Devices are plain
data consumed by the engine's compilation pipeline, which also produces the
resource counts, and interop adapters convert blocks at the boundary. Lists are indicative, not exhaustive.}
\label{fig:arch}
\end{figure}
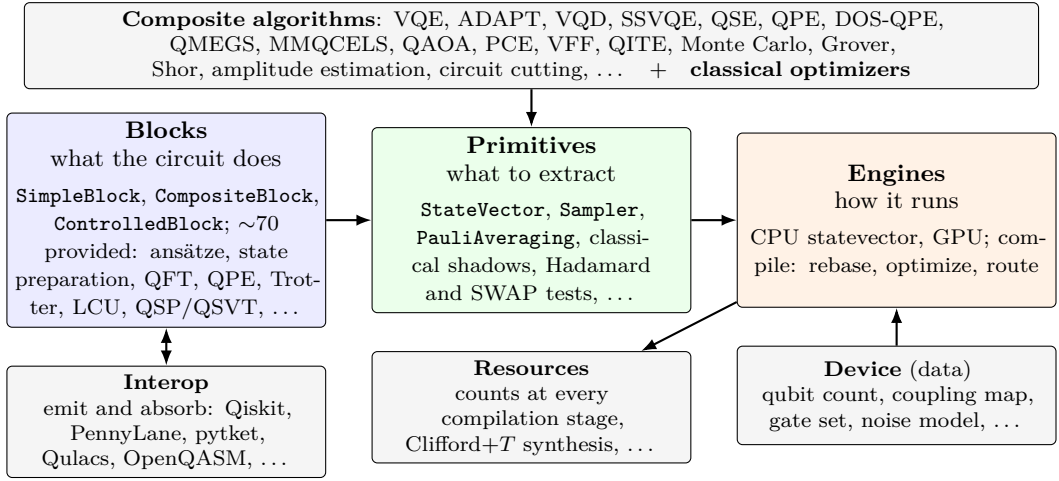

\subsection{Blocks, primitives, engines}
\label{sec:bpe}

\qarp{} separates what a circuit does from what you want out of it, and both
from how it runs. Figure~\ref{fig:arch} maps the layers and
Figure~\ref{fig:hello} shows them on a Bell state.

The unit of circuit construction is the \emph{block}: an object that stands
for a meaningful operation, such as an ansatz layer, a Trotter step, or a
state-preparation routine, rather than a flat list of gates. The abstraction
is inspired by Yao.jl~\cite{luo2020yao}, which builds circuits from composable
blocks in Julia. Blocks carry
symbolic parameters, know their own adjoints, and compose. A
\texttt{CompositeBlock} joins sub-blocks together on their target qubits,
and a \texttt{ControlledBlock} wraps any unitary block with an arbitrary
number of controls. Users who want gate-level control subclass
\texttt{SimpleBlock} and write \texttt{h}, \texttt{cx}, \texttt{rz}, and so
on directly. Users who want higher-level structure assemble the provided
blocks or contribute their own.

A \emph{primitive algorithm} declares the measurement protocol and the
quantity of interest: exact statevector evaluation, shot-based sampling,
expectation estimation by Pauli averaging, classical
shadows~\cite{huang2020shadows}, or basis-rotation grouping of molecular
Hamiltonians~\cite{huggins2021basis}. Overlap and transition-amplitude
protocols are provided in the Hadamard-test~\cite{cleve1998revisited},
SWAP-test~\cite{buhrman2001swap}, and
mirror-test~\cite{kyriienko2020inverse,scali2024qtda} families. The target quantity, whether a
sampling distribution, $\braket{\psi|O|\psi}$, $\braket{\phi|\psi}$, or
$\braket{\phi|O|\psi}$, is inferred from what the user supplies. The
infinite-shot limit has its own value, \texttt{qarp.EXACT}, which the engine
reads as a request for the exact Born distribution. It is a typed enumeration
and not a magic number, so a leaked one raises an error rather than computing a
plausible shot count, and the same experiment moves between idealized and
sampled regimes by changing one argument.

An \emph{engine} is responsible for execution. It takes a list of primitives,
compiles them for a target device when one is given, and runs them under a
parameter binding: a mapping from the circuit's symbolic parameters to numerical values,
which lets one compiled circuit be re-executed at many parameter settings. The
default engine performs fast statevector simulation on CPU, and a GPU engine
is available as an optional extra. Device concerns (gate set, connectivity,
noise) attach to the engine rather than to the primitive, so an algorithm
specification carries over from idealized simulation to a device-constrained,
noisy configuration, with the primitive changed to a sampling one where exact
amplitudes are no longer defined.

The purpose of separating blocks, primitives, and engines is flexibility. A
researcher can swap the measurement strategy under a fixed ansatz, or move the
same algorithm onto a compiled, noisy device, by changing the primitive and the
engine and nothing else. Figure~\ref{fig:swap} does both on the running
example: exact estimation becomes sampled, and then the engine acquires a noise
model. Asking that engine for exact amplitudes raises a typed error at build
time, which is also why a variational loop on a noisy or routed device runs
through a sampling primitive and not through the exact one. The same
freedom applies to the framework itself: an engine is an interface, and
backends can be added or replaced as simulation technology evolves. Composite
algorithms (Section~\ref{sec:algorithms}) are thin orchestrations of
primitives plus classical optimization, which is why new methods from the
literature are typically added to \qarp{} quickly.

\begin{figure}[t]
\begin{lstlisting}
from qarp.blocks import SimpleBlock
from qarp.algorithms import Sampler, StateVector
from qarp.engines import QarpEngine
from qarp.operators import QubitOperator

bell = SimpleBlock(2, name="bell")
bell.h(0)
bell.cx(0, 1)
bell.build()

H = QubitOperator("Z0 Z1") + 0.5 * QubitOperator("X0 X1")

exact = StateVector(ket=bell, operator=H)      # <H> exactly
shots = Sampler(ket=bell, n_shots=4000)        # bitstring distribution

engine = QarpEngine(seed=42)
engine.build([exact, shots])
results = engine.run()
# results[0] -> 1.5 + 0j;  results[1] -> {(0,0): 0.489, (1,1): 0.511}
\end{lstlisting}
\caption{The three layers in one example: a \emph{block} defines the
circuit, \emph{primitives} declare what to extract from it, and the
\emph{engine} runs both against the same state.}
\label{fig:hello}
\end{figure}

\begin{figure}[t]
\begin{lstlisting}
from qarp.algorithms import PauliAveraging
from qarp.devices import NoiseModel

sampled = PauliAveraging(ket=bell, operator=H, n_shots=4000)  # same block, shots
ideal = QarpEngine(seed=42)
ideal.build([exact, sampled])
ideal.run()          # [1.5+0j, 1.5]: exact and sampled estimates agree

noise = NoiseModel.depolarizing(0.05, "2q")                  # on every 2q gate
noisy = QarpEngine(n_qubits=2, noise_model=noise, seed=42)
noisy.build([sampled])
noisy.run()          # [1.3355]: same circuit and protocol, noisy device

noisy.build([exact]) # CapabilityError: amplitudes are undefined under noise
\end{lstlisting}
\caption{Changing one thing at a time. The block and operator of
Figure~\ref{fig:hello} are reused unchanged. Swapping the primitive moves from
exact to sampled estimation, swapping the engine's device moves to noisy
simulation, and asking a noisy engine for exact amplitudes fails with a typed
error at build time rather than returning a plausible number.}
\label{fig:swap}
\end{figure}

\subsection{One set of conventions}
\label{sec:contract}

A framework that composes circuits programmatically depends entirely on its
conventions. \qarp{} keeps a single authoritative conventions document,
versioned with the code; when implementation and document disagree, the
documented convention takes priority and the code is fixed. Three choices are
worth highlighting because each removes a whole class of silent error.

First, qubit~0 is the least-significant bit everywhere~\cite{smith2017ordering}:
in measurement
outcomes, statevector indices, sampler keys, and operator matrices alike.
Operators and states therefore contract directly, with no internal boundary
at which a bit reversal can be forgotten. Other packages' orderings are
handled at the interop boundary, where the conversion is explicitly enforced.

Second, all angles are radians and every rotation follows the
$e^{-i\theta P/2}$ convention. Half-turn inputs, a recurring source of silent
factor-$\pi$ bugs when moving between SDKs, appear nowhere.

Third, global phase is treated as physical. A block's global phase is
unobservable on its own but becomes a relative phase the moment the block sits
under control or inside a linear combination of unitaries; phase estimation
over a synthesized unitary with an uncalibrated phase silently returns shifted
eigenvalues. Every block that can be controlled must therefore implement its
unitary exactly, phase included, and the circuit transformations preserve that
exactness. The corresponding tests compare exact equality rather than equality
up to phase.

We met this during development. A Pauli-block constructor split a complex
coefficient into a global phase and a magnitude, and applied the phase only
when the imaginary part was nonzero, so a negative real coefficient lost its
sign and a block asked for $-Z$ built $+Z$. On its own that block was
indistinguishable from a correct one, and a comparison up to phase passed it.
Under control the dropped sign would have become a relative phase, and every
result computed through it would have been wrong by an amount no convergence
criterion reports. The exact comparison caught it.

\subsection{A compiled core behind a research surface}
\label{sec:core}

All of \qarp{}'s performance-critical machinery, including operator algebra,
fermionic mappings, circuit optimization, routing, and simulation, is
contained in a C++ core bound to Python with
nanobind~\cite{jakob2022nanobind}. The binding is tight: Python block classes
are genuine subclasses of their C++ counterparts, so building a circuit
populates a native buffer directly and nothing is serialized at the language
boundary. Gate parameters remain symbolic inside the core and are substituted
at bind time, which is what makes variational workloads cheap: a parameterized
circuit is compiled once and re-executed across optimizer iterations,
including batched parameter sweeps that never re-enter Python.

For the actual statevector updates, the current release deliberately reuses
existing work: it vendors the well-tested simulation kernels of
Qulacs~\cite{suzuki2021qulacs} and wraps them with a thin interface. The
framework's own contribution is everything around those per-gate kernels, and
it is what separates the two stacks in Section~\ref{sec:bench-statevector}:
gate fusion into dense multi-qubit blocks, an expectation kernel that
contracts a Hamiltonian in one pass per group of terms, a fast path for
structured phase-estimation circuits, batched execution, and the compilation
pipeline. We consider the kernel an exchangeable implementation detail behind
the engine interface, which we expect to revisit as simulation technology
advances.

Two further properties shape everyday use. Circuit optimization passes are
verified to preserve semantics exactly, so users can enable them without
fear of affecting correctness. Furthermore, capability failures are typed:
any request an engine or exporter cannot honor raises one dedicated
exception class carrying the offending operation, so fallback logic in user
code is a single \texttt{except} clause rather than a per-backend taxonomy.
Non-unitary structure (mid-circuit measurement, reset, classically
conditioned blocks) is supported and automatically routed to a trajectory
simulation path, while purely unitary circuits stay on the fast path.

\section{What is available today}
\label{sec:features}

This section surveys the current release. We describe it in the present tense,
but the reader should take the lists as a snapshot of a growing library: each
area is defined by the interfaces of Section~\ref{sec:philosophy} and is built
to be extended.

\subsection{Operators and chemistry}
\label{sec:operators}

\qarp{}'s \texttt{QubitOperator} and \texttt{FermionOperator} keep
OpenFermion's familiar API~\cite{mcclean2020openfermion} over a C++ core that
holds a Pauli string in binary-symplectic form, one pair of 64-bit masks per
64 qubits, following the PauliEngine design~\cite{muller2026pauliengine}. A
product of two strings is then an exclusive-or of the masks with a phase from
four population counts, and a commutation test is two, both independent of how
many qubits the terms act on. Terms are held in an insertion-ordered hash map
that reproduces OpenFermion's iteration order, so results that depend on term
order agree term by term. Section~\ref{sec:bench-operators} measures what the
representation is worth. The operator layer covers
algebra (construction, accumulation, products, commutators, conjugation,
spectra), the standard fermion-to-qubit mappings
(Jordan--Wigner~\cite{jordan1928},
Bravyi--Kitaev~\cite{bravyi2002,seeley2012}, and parity), and measurement
grouping strategies for expectation estimation. Operators can carry symbolic
coefficients through an optional SymEngine backend~\cite{symengine}, so
parameterized generators can be derived once and bound numerically later.

Electronic-structure problems enter as plain NumPy integral tensors, with
active-space reduction and standard model builders (Fermi--Hubbard
\cite{hubbard1963}, transverse-field Ising, and others); a chemistry package
such as pyscf~\cite{sun2018pyscf} is needed only to generate integrals.
Native coupled-cluster excitation generators~\cite{romero2018ucc} feed the
unitary coupled-cluster (UCC) ansatz and the adaptive algorithms. Conversion to and
from OpenFermion objects is provided for interop.
\begin{lstlisting}
from qarp.operators import QubitOperator, FermionOperator
from qarp.operators import JordanWigner, BravyiKitaev

H = QubitOperator("Z0 Z1") + 0.5 * QubitOperator("X0 X1")
excitation = FermionOperator("2^ 0") - FermionOperator("0^ 2")

H_jw = JordanWigner().encode_operator(excitation)   # also BravyiKitaev, Parity
H_bk = BravyiKitaev().encode_operator(excitation)
\end{lstlisting}

\subsection{Blocks}
\label{sec:blocks}

The block library numbers around seventy classes, spanning state
preparation~\cite{bartschi2019dicke,mottonen2005transformation},
ans\"atze~\cite{kandala2017hea,romero2018ucc,krompiec2026seniority,farhi2014qaoa},
algorithmic structure~\cite{coppersmith1994qft,kitaev1995qpe,nielsen2010,lloyd1996,yoshida1990,morales2025product,grover1996,brassard2002amplitude},
and operator-level machinery: exponentials of commuting Pauli sets synthesized
through a shared basis-change Clifford~\cite{aaronson2004stabilizer}, block
encodings for linear combinations of unitaries~\cite{childs2012lcu},
qubitization~\cite{low2019qubitization}, and QSP/QSVT
sequences~\cite{martyn2021grand,gilyen2019qsvt}. Multi-controlled operations
lower without ancillas~\cite{barenco1995}, and symmetry projectors support
symmetry-restored measurements.

Because control wrapping accepts any unitary block and the stack preserves
phases exactly, these pieces compose soundly: a synthesized time-evolution
block can sit under a phase-estimation ladder, or a user-defined ansatz under
an amplitude-amplification reflection, and the result is correct by
construction. A block written by a user enjoys exactly the same treatment as
one distributed with the package.
\begin{lstlisting}
from qarp.blocks import CompositeBlock, ControlledBlock, HEABlock, QFTBlock

ansatz = HEABlock(n_qubits=4, n_layers=2, real=False,
                  linear=True, circular=False, use_cz=False)
qft = QFTBlock(n_qubits=4)

circuit = CompositeBlock([ansatz, qft], n_qubits=4).build()
controlled = ControlledBlock(qft, num_controls=1).build()   # any unitary block
\end{lstlisting}

\subsection{Algorithms}
\label{sec:algorithms}

Composite algorithms orchestrate primitives and classical optimization
behind a uniform interface. Table~\ref{tab:algorithms} lists the current
library. The variational family shares consistent parameter handling, and
the classical side offers SciPy wrappers,
Rotosolve~\cite{ostaszewski2021rotosolve}, SPSA~\cite{spall1992spsa}, and a
gradient-descent family including Adam~\cite{kingma2015adam}, with analytic
gradients available from the engine. Because an algorithm here is a thin
layer over primitives, adding a method from the recent literature can be
achieved rapidly.
\begin{lstlisting}
from qarp.algorithms import VQE
from qarp.blocks import HEABlock
from qarp.optimizers import ScipyOptimizer

ansatz = HEABlock(n_qubits=2, n_layers=2, real=True,
                  linear=True, circular=False, use_cz=False)
vqe = VQE(operator=H, ket=ansatz,            # H from the operator layer above
          optimizer=ScipyOptimizer("COBYLA", options={"maxiter": 500})).build()
vqe.run()
energy = vqe.result.fun                      # -1.5, the exact ground state
\end{lstlisting}

\begin{table}[H]
\centering
\small
\begin{tabular}{@{}lll@{}}
\toprule
Family & Algorithms & References \\
\midrule
Ground state & VQE, ProjectedVQE, AdaptVQE, QITE & \cite{peruzzo2014vqe,seki2022projection,grimsley2019adapt,motta2020qite} \\
Excited states & VQD, SSVQE, AdaptVQD, QSE & \cite{higgott2019vqd,nakanishi2019ssvqe,chan2021adaptvqd,mcclean2017qse} \\
Phase estimation & QPE, DOS-QPE & \cite{kitaev1995qpe,nielsen2010,scali2025dosqpe} \\
Early-FT spectral & QMEGS, MMQCELS & \cite{ding2024qmegs,ding2023mmqcels,marquez2026nmr} \\
Optimization & QAOA, PCE and variants & \cite{farhi2014qaoa,sciorilli2025pce,padin2026pbpce,soloviev2026cpce} \\
Simulation \& sampling & VFF, MonteCarlo & \cite{cirstoiu2020vff,buonaiuto2026qcqmc} \\
Search \& factoring & Grover, Shor, AmplitudeAmplification, AmplitudeEstimation & \cite{grover1996,shor1997,brassard2002amplitude} \\
\bottomrule
\end{tabular}
\caption{Composite algorithms in the current release, all implemented over
the same block/primitive/engine stack.}
\label{tab:algorithms}
\end{table}

\subsection{Devices, compilation, and noise}
\label{sec:devices}

A device in \qarp{} is data: qubit count, coupling architecture, gate set, and
noise model. The engine is responsible for the compilation pipeline that
rebases, routes, and re-targets a circuit for that device, and a standalone
helper exposes the same pipeline without execution. The current router
implements the SABRE heuristic~\cite{li2019sabre} with a portfolio search over
it; routing, like simulation, is defined by an interface, and alternative
routers can be selected or added. Noise models are assembled from standard
channel builders (depolarizing, Pauli, bit-flip, and amplitude damping) and
composed by addition.
\begin{lstlisting}
from qarp.devices import Device, NoiseModel, get_nearest_neighbour_architecture
from qarp.engines import QarpEngine

noise = (NoiseModel.depolarizing(0.01, gate_set="2q")
         + NoiseModel.amplitude_damping(0.02, gate_set="1q"))

device = Device(n_qubits=4,
                architecture=get_nearest_neighbour_architecture(2, 2),
                noise_model=noise.inner)
engine = QarpEngine(device=device, seed=42)
\end{lstlisting}

\subsection{Interoperability}
\label{sec:interop}

\qarp{} treats other SDKs as peers rather than dependencies: no core code path
imports an external SDK, and adapters import lazily inside the call. Circuits
are emitted to Qiskit, PennyLane, pytket, Qulacs, OpenQASM~2 and~3
source text~\cite{cross2022openqasm3}, and QIR, and absorbed from all of these
but QIR. What matters is the failure mode: every exporter declares its
capabilities and validates before converting, so an untranslatable circuit
raises a typed error naming the offending operation rather than producing a
silently wrong export. Symbolic parameters survive the OpenQASM~3 boundary, so
a parameterized circuit stays parameterized on the other side.
\begin{lstlisting}
from qarp.emit import QiskitEmitter
from qarp.absorb import QiskitAbsorber

qc = QiskitEmitter().emit(block.flatten(), block.n_qubits)   # out to Qiskit
back = QiskitAbsorber().absorb(qc)                           # and back again
qasm = block.to_qasm3()                     # or to_qasm2(), to_qir(); QIR is export-only
\end{lstlisting}

\subsection{Cutting and resources}
\label{sec:cutting-resources}

For problems wider than the available register, \qarp{} implements circuit
cutting~\cite{peng2020cutting} by quasi-probability decomposition of
two-qubit gates~\cite{mitarai2021gatecutting}, with an automated search for cut
locations (differential evolution over qubit partitions, minimizing the
number of cuts under subcircuit-size constraints) and a reconstruction layer
that reassembles expectation values from the fragments. A configurable cap on
the number of cuts guards the $6^{k}$ distinct circuit experiments that $k$
cuts require. That count is experiments, not shots: each experiment carries
its own shot allocation and its own random stream, so the total shot cost at a
target precision is the number of experiments times the shots each one needs,
with the variance of the quasi-probability sum on top. A fraction of the
experiments can be run in place of all of them, either as an unbiased uniform
subsample or, when the decomposition's coefficients are skewed, as the
dominant terms.

A resource-estimation layer assesses the cost of a circuit at each stage of
the compilation pipeline, from logical counts through routing overhead to a
target gate set, with an optional synthesis stage that lowers rotations to
Clifford$+T$ using the Ross--Selinger method~\cite{ross2016gridsynth} at a
declared accuracy. Resource-estimation reports follow a simple discipline that
downstream tooling can rely on: unknown quantities are reported as unknown
rather than as zero, and counted quantities are never mixed with modeled ones.

\section{An application end to end}
\label{sec:usecase}

The sections above describe the parts separately. One of the release's
use-case notebooks runs the whole path, from a molecule to a spectrum, twice
on the same system.

Twisting ethylene about its C=C bond drives the low-lying electronic states
through a conical intersection and leaves the ground state degenerate: a
multireference problem on which single-determinant methods fail. Geometries
are rigid rotations of a planar structure, integrals come from
pyscf~\cite{sun2018pyscf} through the plain-tensor interface of
Section~\ref{sec:operators}, and exact diagonalization of each active-space
Hamiltonian is the reference throughout.

At a fixed $90^\circ$ twist, flapping one methylene out of plane sweeps the
molecule through an $S_0/S_1$ intersection. In a CAS(2,2) active space, four
qubits, DOS-QPE~\cite{scali2025dosqpe} phase-estimates $e^{+iH\tau}$ against a
Dicke probe $|D_{4,2}\rangle$, a uniform mixture over the two-electron sector,
so the sampled distribution peaks at the eigenphases of the triplet and the
two lowest singlets at once. Away from the intersection the three are
resolved; at a flap of $68^\circ$ the two singlet peaks merge, which is the
intersection seen directly in the sampled density of states. The same run
exercises a state-preparation block, a composite algorithm, and the sampling
primitive over one engine.
\begin{lstlisting}
import numpy as np
from qarp.algorithms import DOSQPE
from qarp.blocks import TrotterBlock
from qarp.engines import QarpEngine

# TrotterBlock builds exp(-iHt), so a negative time gives U = exp(+iHt)
unitary = TrotterBlock(operator=H_active, n_qubits=4, steps=2,
                       time=-2 * np.pi, order=2).build()

dosqpe = DOSQPE(unitary, n_ancilla=8, hamming_weight=2,   # Dicke probe |D_4,2>
                engine=QarpEngine(seed=7)).build()
dosqpe.run()
spectrum = dosqpe.distribution      # peaks at the eigenphases of that sector
\end{lstlisting}

The second pass takes the full $0^\circ$ to $90^\circ$ torsion in CAS(2,3),
six qubits. A short unitary coupled-cluster VQE supplies a walker basis, and
QC-QMC~\cite{buonaiuto2026qcqmc} refines the ground-state energy by
imaginary-time walker dynamics. Twisted ethylene is a diradical, so the
restricted reference is not unique at the perpendicular geometry; the
active-space spectrum is checked against a pyscf CASCI calculation on the same
orbitals, and at $90^\circ$ the ground state is a triplet with the lowest
singlet a few mHa above it.

Neither pass required code outside the framework, and the notebook runs in the
nightly job of Section~\ref{sec:testing}, so a change that breaks the path from
geometry to spectrum breaks a test.

\section{Correctness methodology}
\label{sec:testing}

The failure mode this section exists to prevent is the one from
Section~\ref{sec:intro}: software that returns a number rather than an error.
A convergence criterion, a variational bound, and a unit test written against
the implementation's own output will all pass a wrong answer. The rule that
does not is an independent oracle, and \qarp{} applies it everywhere a number
is produced: every numerical feature is tested against an analytic value, a
published result, or an OpenFermion or SciPy reference, never against
\qarp{}'s own output. Every test is reviewed against one question: would it
fail if the feature were wrong?

Three structures make that rule executable. First,
the conventions document of Section~\ref{sec:contract} has an executable form:
a reference implementation built from the document by hand, sharing no code
with the simulator, against which the C++ circuit transformations are pinned,
so the kernel, the passes, and the document cross-check one another. Every
gate matrix, argument ordering, and identity the document states is checked
against a NumPy and SciPy oracle constructed from first principles, so a
violated convention fails a test. Second, property-based
tests~\cite{maciver2019hypothesis} search invariants over generated inputs
rather than over cases an author thought of, and every counterexample found is
pinned as a permanent regression case. Third, the compiled-circuit
equivalence machinery of Section~\ref{sec:bench-compilation} applies to the
compiler what the oracle rule applies to the kernels, including negative
controls that confirm a deliberately unrouted circuit is rejected.

At the time of writing the suite holds about 3000 Python test functions in
249 files and 47 C++ test files. CI enforces a total-coverage floor of 94\%
that cannot be reduced, together with 80\% patch coverage on every pull
request. The example notebooks, around sixty at release including a guided
tutorial, are executed nightly, so breaking an example is breaking a test.
This does not establish that the physics is right. It reduces the number of
ways it can be wrong without failing a test. Section~\ref{sec:benchmarks}
extends the same discipline to performance claims, where a check accompanies
every published number.

\section{Benchmarks}
\label{sec:benchmarks}

\subsection{Methodology}
\label{sec:bench-method}

Performance claims about quantum software are fragile. A comparison can be
silently biased by a slow default backend, an easier coupling map, mismatched
gate sets, or simply by a wrong answer returned quickly. The \qarp{}
repository~\cite{openqarp2026} therefore includes the complete benchmark suite
as a public package, with six tracks: operator algebra, statevector
simulation, sampling, end-to-end algorithm optimization, compilation quality,
and code volume. The ground rules are as important as the numbers.

\begin{itemize}
\item \textbf{Every row carries a correctness check.} Each workload reduces
  its result to a canonical physical summary, compared against an
  independent reference: OpenFermion for operators, a plain-NumPy
  statevector implementation, independent of every SDK under test, for
  circuits, and exact diagonalization or CCSD for the code-volume
  implementations. A row whose check fails is not published, and a nightly CI job
  re-runs the checks.
\item \textbf{Competitors run their fast configurations.} Aer is used for
  Qiskit, Lightning for PennyLane, and qsim for Cirq. On the statevector
  track every engine additionally runs the gate fusion its public API
  exposes, at the setting a per-host tuning sweep measured fastest: Aer
  five-qubit blocks from 14 qubits, Qulacs its light optimizer, qsim
  three-qubit blocks, Lightning its transform disabled. Fusion is inside the
  timed region for every stack, \qarp{} included, because a fused block bakes
  in the gate parameters and a parameter sweep pays that cost once per
  parameter set.
\item \textbf{Same problem, same target.} The compilation track hands every
  compiler the identical coupling map and target basis, computes all metrics
  from the compiled gate lists itself, and checks every compiled circuit
  against the coupling map and against the circuit it was compiled from
  (Section~\ref{sec:bench-compilation}).
\item \textbf{Measurement hygiene.} Every measurement runs in a fresh
  subprocess, with medians of at least three repeats, pinned threads,
  per-stack warmup, and identical treatment of setup costs on every stack.
\end{itemize}

The numbers below come from the published tables generated on an arm64 macOS
laptop (September 2026) with \qarp{} 0.1.0, qiskit 2.5.2 with qiskit-aer
0.17.2, pennylane 0.45.1 with lightning 0.45.0, qulacs 0.6.14, qsimcirq
0.22.1, openfermion 1.7.1, and pytket 2.18.1, all on Python 3.13.7; the
code-volume track adds qiskit-nature 0.8.0 with qiskit-algorithms 0.4.0. Each
generated table records host, date, and versions. A third party regenerates
any of them from the release commit with \texttt{python -m benchmarks.run
<track>}, and \texttt{python -m benchmarks.smoke} runs every correctness check
at small sizes, which is also what CI executes nightly so the benchmarks stay
valid.

\subsection{Operator algebra}
\label{sec:bench-operators}

Table~\ref{tab:operators} compares the \qarp{} operator layer with OpenFermion
(the track's correctness oracle), Qiskit's \texttt{SparsePauliOp}, and
PennyLane. \qarp{} runs 10 to 50 times faster than OpenFermion across
construction, accumulation, products, commutators, and all three fermionic
mappings, and the gap widens with system size on the mappings (29 times at 20
modes, 50 at 40). On symbolic coefficients the speedup exceeds two orders of
magnitude. \qarp{} is not the fastest stack on one row: Hermitian conjugation,
where Qiskit's \texttt{SparsePauliOp} is an order of magnitude quicker. These
are the operations that dominate method development: building Hamiltonians,
deriving commutator pools for adaptive ans\"atze, and manipulating
parameterized generators.

The margin comes from the representation of Section~\ref{sec:operators} rather
than from the implementation language. OpenFermion keys a term by a tuple of
(qubit, Pauli) pairs in a Python dictionary, so a product is a merge over two
such tuples and a commutator is two of those merges for every pair of terms,
at a cost that grows with the weight of the terms. \qarp{} keys a term by its
mask pair, so the same product is an exclusive-or and four population counts at
any weight. The spread across the table follows: construction and accumulation,
which are dominated by hashing and by the container, run 5 to 11 times faster
than OpenFermion, while products run 15 to 17 times faster and commutators 35.
The fermionic mappings widen with system size for the same reason, with the
whole operator transformed in a single pass. Symbolic coefficients are a
separate mechanism. The coefficient is a template parameter of the term
container, and the symbolic backend is SymEngine~\cite{symengine}, a C++
computer-algebra library, against OpenFermion's SymPy; that is the 377-fold
row, and an operator with numeric coefficients pays nothing for it. Two rows
bound the argument. Qiskit's \texttt{SparsePauliOp} holds an operator as dense
bit arrays with a coefficient vector, so Hermitian conjugation is one
vectorized pass with no per-term work at all, and no term-wise representation
beats that shape on that operation. At a million terms the cost is allocation
rather than algebra, and the margin over OpenFermion narrows to 1.3.

\begin{table}[H]
\centering
\small
\begin{tabular}{@{}llrrrr@{}}
\toprule
workload & size & OpenFermion & Qiskit & PennyLane & \qarp{} \\
\midrule
construct\_string & 100\,000 terms & 250.2 & 1430 & 156.2 & 22.1 \\
accumulate & 100\,000 terms & 211.0 & 579.4 & 77.5 & 21.7 \\
op\_product & 300 terms & 66.6 & 8.3 & 114.6 & 4.4 \\
commutator & 300 terms & 450.5 & 13.5 & 374.3 & 12.8 \\
hermitian\_conjugated & 100\,000 terms & 15.9 & 0.6 & 24.4 & 5.4 \\
jw\_molecular & 20 modes & 59.5 & n/a & 81.1 & 2.0 \\
jw\_molecular & 40 modes & 453.7 & n/a & 495.2 & 9.2 \\
bk\_molecular & 20 modes & 56.1 & n/a & 191.3 & 2.3 \\
parity\_encode & 16 modes & 100.3 & n/a & 47.3 & 4.1 \\
sparse\_construct & 14 qubits & 1090 & 214.3 & 241.2 & 147.3 \\
symbolic\_algebra & 500 terms & 30\,960 & n/a & n/a & 82.1 \\
memory\_hold (s) & $10^6$ terms & 11.33 & 14.36 & 9.50 & 8.59 \\
\bottomrule
\end{tabular}
\caption{Operator-algebra benchmarks (medians of 3, single-threaded; times in
ms unless noted). Every row agrees with the OpenFermion oracle at $10^{-6}$
relative tolerance. Qiskit cells
include the \texttt{simplify()} needed for an equivalent result, and the
fermionic mappings are n/a for Qiskit.}
\label{tab:operators}
\end{table}

\subsection{Statevector simulation and sampling}
\label{sec:bench-statevector}

Table~\ref{tab:statevector} reports run-phase times for representative circuit
families, with construction excluded identically on every stack. \qarp{}
leads every row of the table except the 14-qubit molecular energy, which
PennyLane's Lightning takes. Two effects account for the margin on the
circuit rows, and both require an engine that receives a whole circuit and a
whole observable. A statevector pass reads and writes the whole state, so a run
costs passes over the state rather than gates: at 20 qubits the state is
16\,MiB and a single-qubit pass costs about half a millisecond on this host,
with the arithmetic units close to idle. \qarp{} folds runs of gates into
dense blocks of at most three qubits before dispatch, which takes the 20-qubit
brickwork, QFT, and Trotter fixtures from 228, 981, and 385 passes to 20, 102,
and 48. The measured cost of a three-qubit block pass times those counts
predicts 38, 192, and 90\,ms against the 40.1, 191.5, and 95.3\,ms in the
table. The molecular rows add a second, independent effect: every Pauli string
sharing a flip mask reads the same pair of amplitudes, so the expectation
kernel groups the Hamiltonian by mask and contracts it in one pass per group.
LiH carries 631 terms in 84 groups and H$_2$O 1086 in 162, so a per-term
engine makes three to seven times more passes over the state. Lightning
arrives at the same pass count by another route: it materializes the
Hamiltonian as a sparse matrix during the build phase and contracts it in one
pass, which takes the H$_2$O row at 4.7\,ms against \qarp{}'s 6.3. That matrix
is 11.6\,MiB at this width and would be about 7\,GB for a 2329-term operator
at 20 qubits, a ceiling the streaming kernel does not have.

\begin{figure}[t]
\centering
\begin{tikzpicture}
\begin{semilogyaxis}[
  width=0.78\linewidth, height=6.4cm,
  xlabel={qubits}, ylabel={run time (ms)},
  xtick={8,12,16,20}, xmin=7, xmax=21,
  legend pos=north west, legend columns=2,
  legend style={font=\footnotesize, cells={anchor=west}, draw=none, fill=none},
  grid=major, grid style={gray!20},
  tick label style={font=\footnotesize},
  label style={font=\small},
]
\addplot[gray, mark=o, dashed] coordinates {(8,1.2) (12,6.2) (16,75.8) (20,4950)};
\addlegendentry{NumPy ref}
\addplot[orange!80!black, mark=diamond*] coordinates {(8,3.8) (12,5.4) (16,16.4) (20,249.7)};
\addlegendentry{Lightning}
\addplot[green!50!black, mark=square*] coordinates {(8,1.2) (12,2.3) (16,7.7) (20,125.0)};
\addlegendentry{Aer}
\addplot[blue!60!black, mark=triangle*] coordinates {(8,0.1) (12,0.4) (16,5.8) (20,116.8)};
\addlegendentry{Qulacs}
\addplot[purple, mark=pentagon*] coordinates {(8,1.3) (12,1.8) (16,4.7) (20,50.7)};
\addlegendentry{qsim (f32)}
\addplot[red, very thick, mark=*] coordinates {(8,0.1) (12,0.3) (16,2.1) (20,40.1)};
\addlegendentry{\qarp{}}
\end{semilogyaxis}
\end{tikzpicture}
\caption{Run-phase time for the brickwork family against register size, from
the rows of Table~\ref{tab:statevector}. The vertical axis is logarithmic, so
the common slope is the shared cost per two added qubits and the vertical
offsets are the constant factors between stacks. The ordering is set by 12
qubits and holds to 20; the 8-qubit points are dispatch-dominated on every
stack, and qsim is single-precision, so its offset is not like-for-like.}
\label{fig:scaling}
\end{figure}
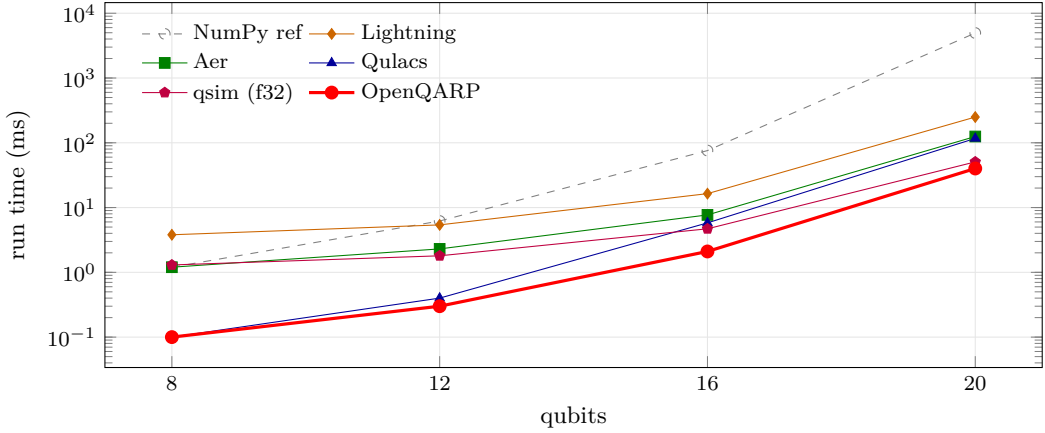

These are not default configurations on the other side: every competitor runs
its own fusion at its tuned setting, and Aer led this table before fusion
landed in \qarp{}. Against the plain-NumPy reference at 20 qubits \qarp{} runs
34 to 123 times faster. Against the fastest competitor the margin is 1.2 to
1.3 times at 20 qubits and wider below 16 qubits, where per-call dispatch
rather than the kernel sets the time. That competitor is usually qsim, whose
float32 wheels halve the bytes moved per pass, a structural advantage no
double-precision engine can claim. Its times are checked at a correspondingly
looser tolerance and are not like-for-like. The ranking is set by 12 qubits
and holds to 20, the largest register in this run: between 16 and 20 qubits
every compiled stack costs about four times as much per two added qubits, and
the offsets in Figure~\ref{fig:scaling} stay in the same order. Sizes from 22
to 26 qubits are a separate opt-in pass of the suite and are not part of this
run. Section~\ref{sec:limitations} states what these numbers rest on and where
we expect them to move. The benchmark suite's thread-scaling axis, not
reproduced here, shows \qarp{}, Aer, and qsim scaling with core count, by 4.5,
4.3, and 3.5 times from one thread to eight, while Qulacs and Lightning stay
flat: the macOS wheels of both link no OpenMP runtime. Peak process memory is
dominated by a fixed import-time cost: \qarp{}'s loaded baseline is about
126\,MiB against 31\,MiB for Qulacs and 76\,MiB for Aer, and well under
Lightning's 357\,MiB and qsim's 192\,MiB. Above that baseline the circuit rows
cost every stack within a few MiB of each other, while Lightning's sparse
expectation holds 133\,MiB at LiH and 695\,MiB at H$_2$O.

\begin{table}[H]
\centering
\small
\begin{tabular}{@{}lrrrrrrr@{}}
\toprule
workload & $n$ & NumPy ref & Aer & Lightning & qsim & Qulacs & \qarp{} \\
\midrule
brickwork & 12 & 6.2 & 2.3 & 5.4 & 1.8 & 0.4 & 0.3 \\
brickwork & 16 & 75.8 & 7.7 & 16.4 & 4.7 & 5.8 & 2.1 \\
brickwork & 20 & 4950 & 125.0 & 249.7 & 50.7 & 116.8 & 40.1 \\
qft & 12 & 8.0 & 3.2 & 8.2 & 3.4 & 0.9 & 0.6 \\
qft & 16 & 125.4 & 13.5 & 36.8 & 14.2 & 19.8 & 8.1 \\
qft & 20 & 8900 & 222.5 & 779.6 & 238.3 & 491.3 & 191.5 \\
trotter\_step & 12 & 4.5 & 2.2 & 5.8 & 2.1 & 0.7\dg & 0.5\dg \\
trotter\_step & 16 & 52.7 & 12.8 & 15.4 & 8.0 & 11.4 & 4.7 \\
trotter\_step & 20 & 3290 & 217.2 & 198.9 & 111.7 & 230.5 & 95.3 \\
qpe\_phase & 14 & 26.4 & 9.2\dg & 14.4 & 6.5 & 6.6 & 2.3 \\
vqe\_molecular & 12 & 69.2 & 7.5 & 2.3 & 32.1 & 3.1 & 1.1 \\
vqe\_molecular & 14 & 343.9 & 43.8 & 4.7 & 140.6 & 19.6 & 6.3 \\
\bottomrule
\end{tabular}
\caption{Statevector run times (ms; medians of 3, single-threaded). Every
stack's state fingerprint agrees with an SDK-independent plain-NumPy
reference at $10^{-9}$ relative tolerance (qsim at $10^{-5}$; see text).
The workloads are random brickwork layers, textbook QFT, Trotter steps of a
transverse-field Ising chain, a QPE-shaped circuit, and one energy evaluation
of an FCI-validated molecular Hamiltonian (12\,q = LiH, 14\,q = H$_2$O) under
a hardware-efficient ansatz. Rows below a millisecond are dominated by
per-call dispatch on every stack and are not a meaningful ranking. A cell
marked $\dag$ had a spread across its three repeats wider than a quarter of
the median.}
\label{tab:statevector}
\end{table}

The sampling track reuses the same circuits and times exactly ``produce the
shots'', checking each stack's output statistics against exact Born values
within six standard deviations of shot noise. Table~\ref{tab:sampler} shows
Qulacs and \qarp{} ahead of the other three stacks by 2 to 20 times, with
Qulacs faster on the brickwork and shot-count rows and \qarp{} faster on the
Trotter and QPE circuits. The two of them separate on how the cost grows with
shots: from 1k to 65k \qarp{} grows about tenfold and Qulacs about fourfold,
while Aer and qsim grow by more than an order of magnitude and Lightning stays
almost flat.

What separates the two leaders is the container, and the two tracks together
separate it from the kernel. The timed call includes whatever each stack's own
sampling API materializes: Qulacs returns a flat list of integers, Lightning
and qsim a shots-by-$n$ bit array, Aer a counts dictionary, and \qarp{} a
dictionary from bitstring to probability. Subtracting the state-preparation
time for the same circuit in Table~\ref{tab:statevector} leaves what sampling
and the container cost. At 12 qubits that is about 2.4\,ms for \qarp{} against
0.2 for Qulacs, which decides the brickwork and shot-count rows on its own; on
the 16-qubit Trotter circuit \qarp{} prepares the state 6.7\,ms faster and
leads the row despite paying more for the container. Part of every gap on this
track is therefore container construction rather than sampling throughput, and
the two stacks that grow fastest with shot count are the two that return a
full result object. The tracks run in separate processes, so the subtraction
is indicative; the suite repeats one cell across both and the spread there is
0.2\,ms.

\begin{table}[H]
\centering
\small
\begin{tabular}{@{}lrrrrrr@{}}
\toprule
workload & $n$ / shots & Aer & Lightning & qsim & Qulacs & \qarp{} \\
\midrule
brickwork\_sample & 8 & 4.1 & 4.4 & 6.3 & 0.2 & 0.3 \\
brickwork\_sample & 12 & 11.6 & 6.2 & 9.9 & 0.6 & 2.7\dg \\
brickwork\_sample & 16 & 22.6 & 19.0 & 14.3 & 6.5 & 7.5 \\
trotter\_sample & 16 & 16.9 & 16.9 & 16.4 & 12.3 & 6.0 \\
qpe\_sample & 14 & 18.1 & 15.4 & 15.3 & 6.8 & 5.8 \\
shots\_axis & 1024 & 4.1 & 6.2 & 3.2 & 0.5 & 0.6 \\
shots\_axis & 8192 & 11.5 & 6.6\dg & 9.9 & 0.6 & 2.9 \\
shots\_axis & 65\,536 & 64.1 & 8.3 & 60.3 & 2.2 & 5.7 \\
\bottomrule
\end{tabular}
\caption{Sampling times (ms; 8192 shots unless noted; medians of 3,
single-threaded). Each stack is timed through its own sampling API including
its native output container, so a richer container costs more to produce. The
\emph{shots} rows sweep the shot count at a fixed 12-qubit brickwork circuit.
A cell marked $\dag$ had a spread across its three repeats wider than a
quarter of the median.}
\label{tab:sampler}
\end{table}

\subsection{End-to-end algorithms}
\label{sec:bench-algorithms}

Single-circuit timings understate what a framework user actually gains, so the
algorithms track measures the wall time of a complete optimizer run. The same
classical optimizer (COBYLA~\cite{powell1994cobyla}, the same SciPy code
object with the same start point and evaluation budget) drives each SDK's
native energy-evaluation path, for VQE on FCI-validated molecular Hamiltonians
and for QAOA MaxCut on a fixed 3-regular graph. Every stack's starting energy
must match the exact reference, its converged energy must land in the same
optimum, and VQE rows must additionally respect the variational bound against
the stored FCI energy.

\begin{table}[H]
\centering
\small
\begin{tabular}{@{}lrrrrrrrr@{}}
\toprule
workload & $n$ & exact ref & Aer & Lightning & qsim & ffsim & Qulacs & \qarp{} \\
\midrule
vqe & 4 & 147.6\ms & 130.4\ms\dg & 195.5\ms & 176.2\ms & 71.9\ms & 52.4\ms & 55.8\ms \\
vqe & 12 & 29.88\,s & 4.01\,s & 1.42\,s & 11.21\,s & 0.88\,s & 1.65\,s & 0.74\,s \\
qaoa & 8 & 70.1\ms & 82.4\ms & 163.9\ms & 166.3\ms & n/a & 25.7\ms & 27.1\ms \\
qaoa & 12 & 220.7\ms & 136.6\ms & 235.2\ms & 255.0\ms & n/a & 51.9\ms & 41.3\ms \\
\bottomrule
\end{tabular}
\caption{End-to-end optimization wall time under an identical COBYLA budget
(medians of 3). \emph{exact ref} is the plain-NumPy reference driven by the
same optimizer. ffsim optimizes its native fermionic LUCJ ansatz from the
Hartree--Fock state, a different problem included to show what a specialized
design point achieves.}
\label{tab:algorithms-bench}
\end{table}

Table~\ref{tab:algorithms-bench} shows \qarp{} fastest on the 12-qubit VQE and
QAOA instances and within 7\% of Qulacs on the two smaller ones. What the VQE
rows measure needs stating precisely. The circuit prepares the Hartree--Fock
determinant, and Brillouin's theorem leaves a real single-rotation ansatz with
no first-order descent direction, so every qubit-side column holds at the
Hartree--Fock energy for the whole budget; the checks enforce this, pinning
the starting energy to Hartree--Fock and requiring the final energy to stay in
that basin. Those rows are therefore a throughput comparison of identical
work, not a race to a converged energy, and the ranking is the per-evaluation
one of Section~\ref{sec:bench-statevector} amplified by the optimizer: the
12-qubit row is 424 energy evaluations of the same LiH Hamiltonian that
Table~\ref{tab:statevector} times once. The specialized fermionic simulator
ffsim~\cite{ffsim} optimizes a different, fermionic ansatz from the same
starting state: its wall time is comparable and it reaches a lower energy
within the same budget, so its column is a contrast in ansatz quality rather
than in simulator speed.

\subsection{Compilation quality}
\label{sec:bench-compilation}

The compilation track measures the quality of the compiled circuit rather than
the speed of the compiler. Every compiler receives the identical coupling map
(line and grid topologies) and the identical target basis, and all metrics are
computed from the compiled gate lists, so depth and two-qubit count mean one
thing across the table. The primary metric is CX-equivalent two-qubit gates,
counting SWAP as three.

Equivalence is checked at three strengths, none of which compares one compiler
against another. Every row compares the state the compiled circuit produces,
with the routing permutation undone, against the uncompiled fixture evolved by
the plain-NumPy reference. At small width the comparison is the full unitary
rather than one column of it, up to global phase and up to an input
relabelling recovered by exhaustive search; where a stack declares an initial
layout, the relabelling that works must equal it, which catches an adapter
reading the wrong field. At every published width an independent equivalence
checker, MQT~QCEC~\cite{burgholzer2021qcec}, decides the compiled circuit
against the original, and the verdict, the checker that produced it, and its
runtime are recorded per cell; all 144 cells of the published table carry such
a certificate. That check fails closed: a probabilistic or
relative-phase-capable verdict, a timeout, and a record that is missing or
stale for this host and commit are all rejected. The nightly job runs a
bounded representative subset of the matrix. Separately, every compiled
circuit is checked against the coupling map, with a negative control
confirming that an unrouted circuit fails that check.

\begin{table}[H]
\centering
\small
\begin{tabular}{@{}lrrrrrrr@{}}
\toprule
workload & $n$ & uncompiled & Qiskit O1 & Qiskit O2 & Qiskit O3 & pytket & \qarp{} \\
\midrule
trotter\_line & 12 & 132 & 318 & 315 & 315 & 471 & 312 \\
trotter\_line & 14 & 182 & 443 & 437 & 437 & 491 & 434 \\
trotter\_grid & 12 & 132 & 240 & 237 & 234 & 276 & 234 \\
trotter\_grid & 14 & 182 & 326 & 323 & 308 & 422 & 314 \\
hea\_line & 12 & 36 & 141 & 141 & 141 & 147 & 135 \\
hea\_line & 14 & 42 & 165 & 165 & 165 & 153 & 159 \\
qv\_line & 12 & 48 & 120 & 117 & 117 & 156 & 117 \\
qv\_line & 14 & 56 & 155 & 150 & 150 & 201 & 155 \\
qv\_grid & 12 & 48 & 75 & 69 & 66 & 84 & 66 \\
qv\_grid & 14 & 56 & 92 & 90 & 93 & 108 & 92 \\
mqt\_line & 12 & 150 & 420 & 417 & 420 & 468 & 411 \\
mqt\_line & 14 & 203 & 584 & 587 & 584 & 491 & 575 \\
mqt\_grid & 12 & 150 & 285 & 270 & 273 & 264 & 261 \\
mqt\_grid & 14 & 203 & 407 & 392 & 401 & 413 & 386 \\
\bottomrule
\end{tabular}
\caption{CX-equivalent two-qubit gate counts (SWAP $=$ 3 CX; lower is
better) for identical coupling maps and target basis. \emph{uncompiled} is
the fixture before routing. Workloads: an all-pairs $ZZ$+$X$ layer
(routing-intensive), a hardware-efficient ansatz,
quantum-volume-like random pairings~\cite{cross2019qv}, and the MQT~Bench
QFT~\cite{quetschlich2023mqtbench}, each routed onto a line and a grid.
Shown are the two largest sizes of every workload that needs routing; the
published table also holds an $n=8$ step and a grid-native ansatz on which
every stack ties.}
\label{tab:compilation}
\end{table}

In Table~\ref{tab:compilation}, \qarp{} produces the best or joint-best
circuit on 9 of 14 rows, is never worse than Qiskit at optimization level 1,
and matches or beats level 3 on 12 of 14 rows; pytket is ahead on two line
instances. At $n=8$, omitted from the table, the stacks tie on the two
Trotter workloads and pytket or Qiskit level~3 wins the remaining five by 7
to 19\%. Compile time, deliberately secondary in this track, is 7 to
46\,ms for \qarp{} on these instances: 2.3 to 9 times slower than Qiskit at
level~1, at most twice level~3's time, and 21 to 81 times faster than pytket.

Both results follow from the router. \qarp{} runs the same SABRE
heuristic~\cite{li2019sabre} Qiskit does, at the same published constants,
which is why the two columns track each other row by row, and it differs in
two places. A layout pass runs first and searches for an embedding of the
circuit's interaction graph into the coupling map; where one exists it needs no
SWAP at all, and no amount of local refinement can improve on it. Where none
exists, the heuristic runs as a best-of-$N$ portfolio over independent seeds
with randomized tie-breaking, 32 trials at the circuit sizes here, each refined
by rounds of forward and reverse traversal, keeping the lowest SWAP count. The
first trial reproduces a single seeded run, so the portfolio cannot return a
worse circuit than one shot of the same heuristic would. The compile time is
the same thing seen from the other side: running the heuristic 32 times is what
costs several times a single Qiskit pass, and this track counts that as the
right trade, since a compiler may take longer if the circuit it emits is
cheaper to run.

\subsection{Code volume}
\label{sec:bench-volume}

The five tracks above measure what the machine does. A sixth measures what the
user writes. Six algorithms --- VQE, SS-VQE, ADAPT-VQE, shot-based VQE, QPE,
and QAOA --- are implemented once per stack across \qarp{} and five
competitors, and lines are counted only after every implementation returns the
same number. The primitive stacks hand-write the same ansatz in the same order
and agree to $10^{-8}$; the two frameworks bring their own excitation ordering
and span a slightly different variational manifold, so they are held to
chemical accuracy. Mutual agreement cannot catch an error that every stack
shares, so each row also carries an independent oracle: exact diagonalization
for the excited-state and shot-based rows, a closed form for QAOA, an analytic
phase for QPE, and CCSD for VQE.

The metric is physical lines excluding blanks, comments, and docstrings, with
imports counted on every side and every file passed through the same formatter
at the same line length. Each competitor gets the best its own ecosystem
offers: OpenFermion's mappings and PySCF driver on the primitive stacks, each
stack's own native gate for a Pauli exponential, and each framework's own
high-level path. \qarp{}'s analytic gradients are switched off, so every side
is gradient-free or finite-difference and \qarp{} gets no win the others would
have to hand-write. Machinery shared between a stack's own scripts lives in
one module per stack, and Table~\ref{tab:volume} charges each script only the
functions it reaches.

\begin{table}[H]
\centering
\small
\begin{tabular}{@{}lrrrr@{\hspace{1.4em}}crr@{}}
\toprule
& & \multicolumn{4}{c}{primitive stacks} & \multicolumn{2}{c}{frameworks} \\
\cmidrule(lr){3-6} \cmidrule(l){7-8}
algorithm & \qarp{} & Qulacs & Cirq & Qiskit & reduction & PennyLane & qiskit-nature \\
\midrule
VQE & 30 & 88 & 92 & 83 & 64--67\% & 32 & 32 \\
SS-VQE & 32 & 75 & 81 & 74 & 57--60\% & 31 & 42 \\
ADAPT-VQE & 34 & 117 & 129 & 112 & 70--74\% & 32 & 32 \\
VQE-shots & 30 & 129 & 134 & 128 & 77--78\% & 28 & 34 \\
QPE & 24 & 67 & 44 & 32 & 25--64\% & 23 & 20 \\
QAOA & 18 & 38 & 40 & 32 & 44--55\% & 22 & 23 \\
\midrule
all six, amortized & 168 & 335 & 328 & 295 & 43--50\% & 168 & 183 \\
\bottomrule
\end{tabular}
\caption{Lines of code for the same six algorithms, counted after every
implementation returns the same number. Qiskit appears twice, once raw
(\texttt{SparsePauliOp}, \texttt{PauliEvolutionGate}, \texttt{Statevector})
and once as qiskit-nature with qiskit-algorithms, so the pair isolates what an
algorithm library is worth with nothing else changing. A primitive-stack cell
is the script plus the functions it reaches in that stack's shared module;
\emph{reduction} is the range of $1-$\,\qarp{}/stack across the three. The
final row counts each shared module once, the cost of building a codebase
rather than one script.}
\label{tab:volume}
\end{table}

The two halves of the table answer different questions. Against a primitive
stack the reduction is 25 to 78\%, and the spread tracks how much plumbing the
algorithm needs. Shot-based VQE is the widest row at 77 to 78\%: measurement
grouping, basis rotation, sampling, and an estimator are the user's to write,
and adding them cost the three shared modules 33 to 39 lines each. QPE against
raw Qiskit is the narrowest at 25\%, because Qiskit ships both a QFT and a
controlled Pauli exponential and little is left to assemble. Against a
framework the result is a wash. PennyLane is shorter than \qarp{} on four of
the six rows and qiskit-nature on two, and \qarp{} takes QAOA at 18 lines
against 22 and 23. The 70\% figure belongs to the primitive-stack half of the
table.

Read together the two halves locate the package. A primitive stack cannot
reach these line counts, because there is no algorithm layer and the plumbing
belongs to the user; a framework does reach them, and amortized over all six
algorithms \qarp{} and PennyLane tie at 168 lines against 295 to 335 for the
primitive stacks. \qarp{} is in both columns at once, which is what the
architecture of Section~\ref{sec:bpe} is for. The one place the two halves
touch is qiskit-nature's SS-VQE cell, its worst at 42 lines, where the
framework does not ship the algorithm and the loop is hand-written over a
generic ansatz.

Writing the competitor implementations is also what the conventions of
Section~\ref{sec:contract} are for. Each of the following produces a wrong
number rather than an error. Qulacs rotations are $e^{+i\theta P/2}$ where
\qarp{}'s and Cirq's are $e^{-i\theta P/2}$, and Qiskit's evolution gate
carries no half-angle, so a wrong guess mirrors the ansatz. Cirq indexes basis
states big-endian where Qulacs and Qiskit are little-endian. OpenFermion
interleaves spin orbitals where qiskit-nature blocks them, so the same
spectrum attaches to different basis states. The sharpest case is QPE: a
Hamiltonian's constant term is a global phase on $U$ and is invisible to every
expectation value in the other five rows, but QPE controls $U$, and a
controlled global phase is a relative phase on the ancilla, so dropping it
shifts every estimated eigenphase by exactly the amount that looks plausible.
Each stack needs its own spelling of that fix. One performance hazard belongs
here too, because it is not a property of hand-written code: Qiskit's evolution
gate re-synthesizes on every parameter bind, at 20.4\,s per energy evaluation
and 17 hours for one VQE, and transpiling once to a fixed gate set first costs
one line and takes it to 117\,ms. qiskit-nature's own UCCSD carries the same
cost at 5.6\,s per objective call. Both Qiskit columns include that line, and
neither finishes without it.

A line count says nothing about what a stack can do. The suite keeps a
capability record alongside the tables, resolved from each stack's live API
rather than from its documentation, and it stays in the repository
(\texttt{benchmarks/code\_volume/capabilities.py}) rather than being
reproduced here. Three properties of it argue against printing it in a paper:
the list of capabilities is ours, a resolved symbol grades nothing about the
implementation behind it, and every stack in the comparison ships new methods
faster than a paper is revised. One reading of it survives those caveats.
Infrastructure is common to the field --- noise models, mid-circuit
measurement, transpilation, and fermionic mappings are in essentially every
stack --- so the algorithm catalog is where these stacks differ, and the
spectral and excited-state end of it in particular. Of the methods in
Table~\ref{tab:algorithms}, SS-VQE, quantum subspace expansion, DOS-QPE, and
QMEGS/MMQCELS are shipped by none of the other five; where they exist at all
it is as the code accompanying the original publication. SS-VQE is also where
qiskit-nature posts its worst cell in Table~\ref{tab:volume}, which is the one
place the two measurements meet.

\subsection{Summary of the benchmarks}

Taken together, the six tracks support a simple conclusion. Operator-heavy
work, such as Hamiltonian engineering and adaptive ansatz pools, gains one to
two orders of magnitude with \qarp{} (Table~\ref{tab:operators}).
Simulation-heavy work runs at or near the front of the compared stacks:
\qarp{} leads every row of Table~\ref{tab:statevector} but the largest
molecular energy, splits the end-to-end rows with Qulacs
(Table~\ref{tab:algorithms-bench}), and splits the sampling track with Qulacs
as well (Table~\ref{tab:sampler}), subject to the caveats in
Section~\ref{sec:limitations}. Hardware-bound work gets a compiler that is
competitive with mature stacks at interactive speeds
(Table~\ref{tab:compilation}). Writing the application takes about half the
lines a primitive stack needs and about as many as a framework
(Table~\ref{tab:volume}). Every number above rests on independently
checked physics, which any reader can re-verify with the distributed
benchmarks. Speed is the perishable result here, since any release can change
it. The durable one is that none of it was traded for the architecture: the
composition, the conventions, and the checks cost no performance, and on the
simulation rows they produced it.

\subsection{Limitations}
\label{sec:limitations}

The preceding tables identify what the current release does not do well, and
we list it here explicitly. The statevector results come from one host, an
Apple-silicon laptop with large caches, unified memory, and no AVX. \qarp{}'s
block kernel is scalar C++ where qsim's and Aer's are vectorized, so on an x86
node with wide SIMD we would expect qsim to match or pass \qarp{} on the
circuit rows and Aer's gap to narrow; the energy rows, which turn on a pass
count rather than on a kernel, should hold. The circuit fixtures are dense and
local, which is the best case for fusion: a circuit of long-range two-qubit
gates on random pairs would fuse poorly on every stack, and those rows would
converge toward the per-kernel ratios. The block kernel rather than the memory
system is what currently limits \qarp{} on these fixtures, so a vectorized
version would take a further factor of one and a half to two off the circuit
rows. These measurements are a starting point rather than a settled result:
characterizing the package's performance properly needs a broader campaign
across hosts and architectures, and that work is ongoing. \qarp{}'s loaded
process footprint is about 126\,MiB against 31\,MiB for Qulacs, an import-time
cost that is significant for many short-lived processes but not for long
simulations. The compiler produces competitive circuits but takes several
times longer than Qiskit to do so, and its pass set is narrower than a
dedicated compiler's. There is no autodiff or machine-learning interface and
no pulse-level control, both of which PennyLane offers. GPU execution is
available as an optional engine but is not part of the published tables. QIR
is export-only, and there is no hardware
execution engine: running on a device means emitting a Qiskit or pytket
circuit and bringing the counts back through the user's own code. Finally,
binary wheels cover glibc Linux (x86-64 and aarch64), macOS (Apple silicon and
Intel), and Windows (x86-64); other platforms, musl-based Linux included,
build the C++ core from source.

\section{Agent-assisted development}
\label{sec:agents}

We expect a growing share of contributions to open-source scientific software
to be drafted with coding agents, and \qarp{}'s contribution process is built
for that. A standard or structural contribution starts as a plan file, copied
from a template in \texttt{docs/contributions/}, that states the scope as an
explicit list of files, the design, an API sketch, a test plan naming the
independent oracles, and a deviations log. The plan opens a draft pull request
and is approved by a reviewer before implementation begins; deviations from
the approved version are recorded as they happen. The tier sets how much of
this applies: a typo needs no plan, a new block type needs one. A human
reviewer approves at both stages.

A contribution must also conform to the framework's conventions.
\texttt{qarp\_conventions.md} governs endianness, angle and phase conventions,
block and engine behavior, and the resource-vector rules of
Section~\ref{sec:cutting-resources}, and it takes priority over both the code
and any plan that disagrees with it. It also has an executable form:
\texttt{tests/test\_conventions\_gates.py} checks every gate matrix, argument
ordering and identity the document states against a NumPy and SciPy oracle
built from first principles. A root \texttt{AGENTS.md} carries the ground
rules and points at the conventions document, and \texttt{CLAUDE.md} is a
symbolic link to it.

The architecture helps as much as the rules do. A contributor who adds a
block, a primitive, or an engine touches one axis and inherits the rest, so
the unit of work is small enough to hold in view at once and its correctness
is decided by the oracles. This helps a new human contributor and is a
precondition for an agent.

Past the plan, a contribution meets the checks of Section~\ref{sec:testing}
unchanged: an independent oracle for every numerical feature, coverage floors
on every pull request, property-based tests over generated inputs, executed
example notebooks, and the compiled-circuit certification of
Section~\ref{sec:bench-compilation}. An agent that reads the conventions
document and follows the plan template meets the same bar as anyone else, and
is refused by the same tests when it does not. \qarp{} itself is developed
under this process, with Claude Code~\cite{anthropic2026claudecode} and the
Codex CLI~\cite{openai2026codex}.

\section{Related work}
\label{sec:related}

Qiskit~\cite{javadiabhari2024qiskit}, Cirq~\cite{cirq2024}, and
PennyLane~\cite{bergholm2018pennylane} are circuit-centric SDKs with large
ecosystems. \qarp{} differs in making the composable, phase-exact block its
primary abstraction, and in shipping the algorithm, compilation, and resource
layers inside one consistently-conventioned package rather than across
satellite projects. pytket~\cite{sivarajah2020tket} is a dedicated compiler
with SDK front-ends. \qarp{}'s compiler is narrower in scope and produces
circuits of comparable cost on the shared-target benchmark of
Section~\ref{sec:bench-compilation}. Among the high-performance simulators,
\qarp{} currently builds on Qulacs'~\cite{suzuki2021qulacs} kernels rather than
competing with them, and its engine interface leaves room for
qsim~\cite{qsim2020} and for whatever supersedes both.
OpenFermion~\cite{mcclean2020openfermion} is the de-facto operator API for
quantum chemistry, and \qarp{} keeps that API while moving the implementation
to C++, using OpenFermion itself as the correctness oracle for the
replacement.

Tangelo~\cite{senicourt2022tangelo} and InQuanto~\cite{inquanto}, open-source
and commercial respectively, are end-to-end chemistry platforms built over
general SDKs and simulators, following the workflow from molecule to energy.
\qarp{} covers less of that pipeline, taking integrals from an external
package, and more of the algorithmic space, with spectral, optimization, and
encoding methods beside the chemistry ones. Recent standalone work on fast
Pauli algebra, such as PauliEngine~\cite{muller2026pauliengine}, addresses the
same bottleneck as a library, where in \qarp{} the equivalent kernels are one
component of an integrated stack. At the specialized end of the design space
sits ffsim~\cite{ffsim}, a simulator restricted to fermionic circuits that
outruns every general-purpose stack on its native ansatz.

\section{Conclusion and availability}
\label{sec:availability}

\qarp{}'s design choices concern interfaces and evidence, not particular
algorithms.
Circuits, measurements, and execution are kept separate so that each can
change. One set of conventions makes independently written components compose
correctly, and no numerical claim, in the test suite or in the benchmarks,
stands without an independent check. The particular algorithms, kernels, and
compilation heuristics in the current release are the present state of an
evolving library, and the architecture is designed so that replacing any of
them is an update, not a rewrite. Our hope is that this makes \qarp{} a
natural place to implement methods that do not exist yet.

\qarp{} is developed openly on GitHub at
\url{https://github.com/OpenQARP/openqarp}~\cite{openqarp2026}, where the
source code, issue tracker, and release history are hosted, with documentation
at \url{https://openqarp.github.io/openqarp/}. The version described here is
0.1.0, the first public release. It is distributed under the Apache License
2.0 and supports Python 3.11 to 3.14. \texttt{pip install openqarp} installs a
binary wheel (Linux, macOS, and Windows; CPython 3.11 to 3.14); \texttt{pip
install .} from a checkout builds the C++ core locally, which needs a C++20
compiler and CMake~$\geq 3.20$ and takes a few minutes. Optional extras add
GPU execution, SDK integrations, and the benchmark suite. The distribution
includes full documentation, around sixty executable example notebooks, the
conventions document, and the complete benchmark suite with one-command
regeneration of every table in this paper.

\section*{Data availability}
The benchmark suite, its inputs, the FCI-validated molecular Hamiltonians,
and the generated tables behind Tables~\ref{tab:operators}
to~\ref{tab:volume} are in the repository under \texttt{benchmarks/},
and every table is regenerated as described in
Section~\ref{sec:bench-method}.

\begin{acknowledgments}
\qarp{} is developed and maintained by the Fujitsu Research of Europe team.
We thank the developers of the open-source packages \qarp{} builds on and
benchmarks against, in particular Qulacs, whose simulation kernels the
current release vendors, and OpenFermion, whose operator API \qarp{}
preserves, as well as the communities behind NumPy~\cite{harris2020numpy},
SciPy~\cite{virtanen2020scipy}, SymPy~\cite{meurer2017sympy}, and
Hypothesis~\cite{maciver2019hypothesis}.
\end{acknowledgments}

\bibliographystyle{apsrev4-2}
\bibliography{main}

\end{document}